\documentclass[useAMS,usenatbib]{mn2e}
\input epsf
\usepackage[usenames,dvips]{color}
\usepackage{graphicx}
\usepackage{amsmath}
\usepackage{multirow}
\usepackage{amssymb} 
\title{Investigation of CTA 1 with Suzaku Observation}
\author[Lupin C. C. Lin et al.]{Lupin C. C. Lin$^{1,2}$\thanks{E-mail:
lupin@crab0.astr.nthu.edu.tw}, Jumpei Takata$^{3}$, Albert K. H. Kong$^{4,8}$, C. Y. Hui$^{5}$, Teruaki Enoto$^{6}$,  
\newauthor
H. K. Chang$^{4}$, Regina H. H. Huang$^{4}$, J. S. Liang$^{4}$, Shinpei Shibata$^{7}$ and C. Y. Hwang$^{2}$\\
$^1$General Education Center, China Medical University, Taichung 40402,  Taiwan\\
$^2$Graduate Institute of Astronomy, National Central University, Jhongli 32001, Taiwan\\
$^3$Department of Physics, The University of Hong Kong, Hong Kong, PRC\\
$^4$Institute of Astronomy and Department of Physics, National Tsing Hua University, Hsinchu 30013, Taiwan\\
$^5$Department of Astronomy and Space Science, Chungnam National University, Daejeon, South Korea\\
$^6$Cosmic Radiation Laboratory, the Institute of Physical and Chemical Research, 2-1 Hirosawa, Wako, Saitama 351-0198, Japan\\
$^7$Department of Physics, Yamagata University, Yamagata 990-8560, Japan\\
$^8$Golden Jade Fellow of Kenda Foundation, Taiwan}

\begin{document}

\date{Mar 2012; ??? 2012}
\pubyear{2012}
\maketitle

\begin{abstract}
We report on an 105 ks {\it Suzaku} observation of the supernova remnant CTA 1 (G119.5+10.2). 
The {\it Suzaku} soft X-ray observation was carried out with both timing mode and imaging mode. 
A $\sim 10'$ extended feature, which is interpreted as a bow-shock component of the pulsar wind nebula (PWN), is revealed in this deep observation for the first time. 
The nebular spectrum can be modelled by a power-law with a photon index of $\sim 1.8$ which suggests a slow synchrotron cooling scenario. 
The photon index is approximately constant across this extended feature.
We compare and discuss our observations of this complex nebula with previous X-ray investigations.
We do not obtain any significant pulsation from the central pulsar in the soft (0.2-12 keV) and hard (10-60 keV) X-ray data.
The non-detection is mainly due to the loss of the precise imaging ability to accurately determine the source contribution.
The spectra of XIS and HXD can be directly connected without a significant spectral break according to our analysis.
Future observations of {\it NuSTAR} and {\it Astro-H} would be able to resolve the contamination and provide an accurate hard X-ray measurement of CTA 1.
\end{abstract}
\section{Introduction}
CTA 1 (G119.5+10.2) is a composite supernova remnant (SNR) with centrally bright X-ray emission \citep{SSS95,Sla97}.
This SNR has a radio shell with a radius of $\sim 54'$, and its kinematic distance estimated from the associated H$_{\rm{\uppercase\expandafter{\romannumeral 1}}}$ emission is $1.4\pm 0.3$~kpc \citep{Pin93}. The X-rays from CTA 1 comprise both thermal and non-thermal emission \citep{SSS95}.
While the outer regions mainly consist of shock-heated thermal emission, the non-thermal component dominates the central emission, which has led to the speculation of a hidden pulsar in this SNR \citep{Sla97}.
In addition to the radio and X-ray observations, the launch of {\it Compton Gamma-Ray Observatory} (CGRO) provided a further insight on the nature of CTA~1.
{\it Energetic Gamma Ray Experiment Telescope} (EGRET) onboard CGRO has revealed a $\gamma$-ray source, 2EG J0008+7307 \citep{Thp95}, in the 2nd EGRET catalogue (or 3EG J0010+7309 in the 3th EGRET catalogue; \citealt{Har99}), which was suggested to be associated with CTA 1 in view of the positional coincidence \citep{Bra98}.
The lack of variability of this extended $\gamma$-ray source as well as its spectral resemblance to a typical $\gamma$-ray pulsar further suggested that a neutron star should reside in this SNR. 
Nevertheless, the limited photon statistics obtained by EGRET did not allow any meaningful pulsation search in the MeV$-$GeV regime.

Observations with {\it ROSAT} have revealed an X-ray point source, RX~J0007.0+7302, at the center of CTA~1 \citep{SSS95}. 
It was also identified as the X-ray counterpart of 2EG~J0008+7307 / 3EG J0010+7309 \citep{Bra98}. 
This point source was identified as a neutron star candidate with the high $\gamma$-ray-to-X-ray and X-ray-to-optical flux ratios as well as its X-ray spectral behaviours \citep{Sla97}, which were found to be similar to those of the only one radio-quiet $\gamma$-ray pulsar known at that time --- Geminga \citep{Bra98,HGCH2004}.
Utilizing the high resolution imaging capability of {\it Chandra}, \citet{HGCH2004} have uncovered a torus$+$jet feature, which resembles the morphology of a pulsar wind nebula (PWN) as seen in other bright $\gamma$-ray pulsars. 
Furthermore, the sub-arcsecond angular resolution of the {\it Chandra} image provided a precise position, which can facilitate multi-wavelength investigations. 
To identify the neutron star nature unambiguously, several studies attempted to search for the spin period of
RX~J0007.0+7302 \citep{BL96,NS97,Sla2004,HGCH2004,LC2005} in different energy bands. 
Nevertheless, no conclusive evidence for pulsations in radio/X-ray was yielded in these previous studies.

With the unprecedented sensitivity of the {\it Large Area Telescope} (LAT) onboard {\it Fermi Gamma-ray Space Telescope}, a $\gamma$-ray pulsar, LAT PSR J0007+7303, was eventually detected in CTA~1 \citep{Abdo2008}. 
This source has a spin-period of $\sim 315.86$~ms and was the first radio-quiet $\gamma$-ray pulsar detected by LAT. 
Its spin parameters imply a characteristic age of $\sim 14000$~yr, which is consistent with the kinematic age of CTA~1 ($\sim 10000-20000$~yr; \citealt{Sla97}). 
This strongly suggests an intrinsic association between the pulsar and SNR. 
With the discoveries of more radio-quiet $\gamma$-ray pulsars \citep{Abdo2009,Saz2010}, we can now examine them as a unique class. 
It has been speculated that the main distinction between radio-quiet and radio-loud $\gamma$-ray pulsars is simply a geometrical effect. 
In order to further investigate their emitting properties, X-ray observations are deeply required.

Although the periodic signals of these $\gamma$-ray pulsars can be significantly detected, Geminga is the only one radio-quiet
$\gamma$-ray pulsar with a detected X-ray pulsation before 2009 \citep{HH92}.
In order to investigate the possible origin of the pulsed X-rays from the radio-quiet $\gamma$-ray pulsars as a whole class,
searches for the possible pulsations from these pulsars have been conveyed with dedicated X-ray observations, including PSR J0007+7303.
With a $\sim120$~ks observation with {\it XMM-Newton} (hereafter {\it XMM}), the X-ray pulsation of PSR J0007+7303 have been independently detected by \citet{Lin2010} and \citet{Car2010}. 
Apart from the X-ray pulsation, this deep {\it XMM} observation also reveals the nebular emission extending for $\sim 2'$ in south-east direction away from the pulsar (see Fig.~1 in \citealt{Car2010}). Together with the torus and jet feature found by \citet{HGCH2004}, all previous X-ray investigations 
suggest LAT PSR J0007+7303 has a complex PWN structure. 
To continue the investigation, we have observed this interesting object with {\it Suzaku}, which enables a detailed analysis in both soft and hard X-ray bands. 
In this paper, we report the results obtained from analysing the {\it Suzaku} data.

\section{Observations}
{\it Suzaku} \citep{Mit2007} was built with two main payloads, five nested conical thin-foil grazing incidence telescopes (XRT; \citealt{Ser2007}) with the effective energy range of 0.2-12 keV and a hard X-ray detector (HXD; \citealt{Taka2007}) with the effective energy range of 10-600 keV.
There are two major detectors for spectroscopy (X-ray spectrometer and X-ray imaging spectrometer) in {\it Suzaku} and each of them is set at the focus of an XRT.
Our observational results on the temporal and spectral analyses mainly come from the X-ray imaging spectrometer (XIS; \citealt{Koyama2007}) and the HXD.
The XIS consists of three front-illuminated (FI: XIS0, XIS2, XIS3) and one back-illuminated (BI: XIS1) CCD detectors, but only the FI chips provide the observations in timing/P-SUM mode.
Because the time resolution of HXD is 61~$\mu$s and that of XIS in the timing mode is $\sim 7.8$~ms, our data is precise enough for the periodicity search to a radio-quiet $\gamma$-ray pulsar with the spin period in tens or hundreds of milliseconds.
However, two of the XIS units with FI chips, XIS0 and XIS2 suffered a catastrophic damage on 2006 November and 2009 June respectively.
Since no useful data can be obtained with XIS2 and a part of the segment in XIS0 is lost, only the observation of XIS3 unit can be used for temporal analysis.
These events seriously decrease the probability to obtain the periodic signals from the faint sources in the soft X-ray band.
There are also two independent detector systems used for the observations of HXD.
The positive-intrinsic-negative (PIN) silicon diodes are sensitive below $\sim 60$~keV in the FOV of $34'\times 34'$, and the Gadolinium Silicate/Bismuth Germanate crystals (GSO/BGO) phoswich counters can detect photons above $\sim 30$~keV in the FOV of $4^{\circ}.5 \times 4^{\circ}.5$.
Because the data obtained from HXD have no imaging capability, we only did the analysis of the HXD-PIN data to avoid the  contamination caused by the large background size of the HXD-GSO, which is much larger than the size of CTA~1.
 
Our observation for CTA 1 was carried out on 2010 January 8 with total exposures of $\sim 105$~ks in the soft X-ray band and $\sim 59$~ks in the hard X-ray band, and it was aimed at the HXD nominal position to RX J0007.0+7302 with (J2000) R.A.=$00^h07^m02^s.2$, decl.=$+73^{\circ}03'07''$ \citep{Sla2004}.
The XIS0 and XIS1 data of {\it Suzaku} were observed in the normal mode without window option, while the {\it Suzaku}/XIS3 data observed within 128 rows were compressed into one dimension to attain a higher time resolution.
All the data reduction and the spectral analyses were performed using XSELECT (ver.~2.4a), FTOOLS (ver.~6.9) and XSPEC (ver.~12.6.0) of HEASOFT (ver.~6.9) with the latest {\it Suzaku}/XIS calibration files (20101108) and {\it Suzaku}/HXD calibration files (20101202).

\begin{figure*}
\begin{center}\includegraphics[width=18.0cm]{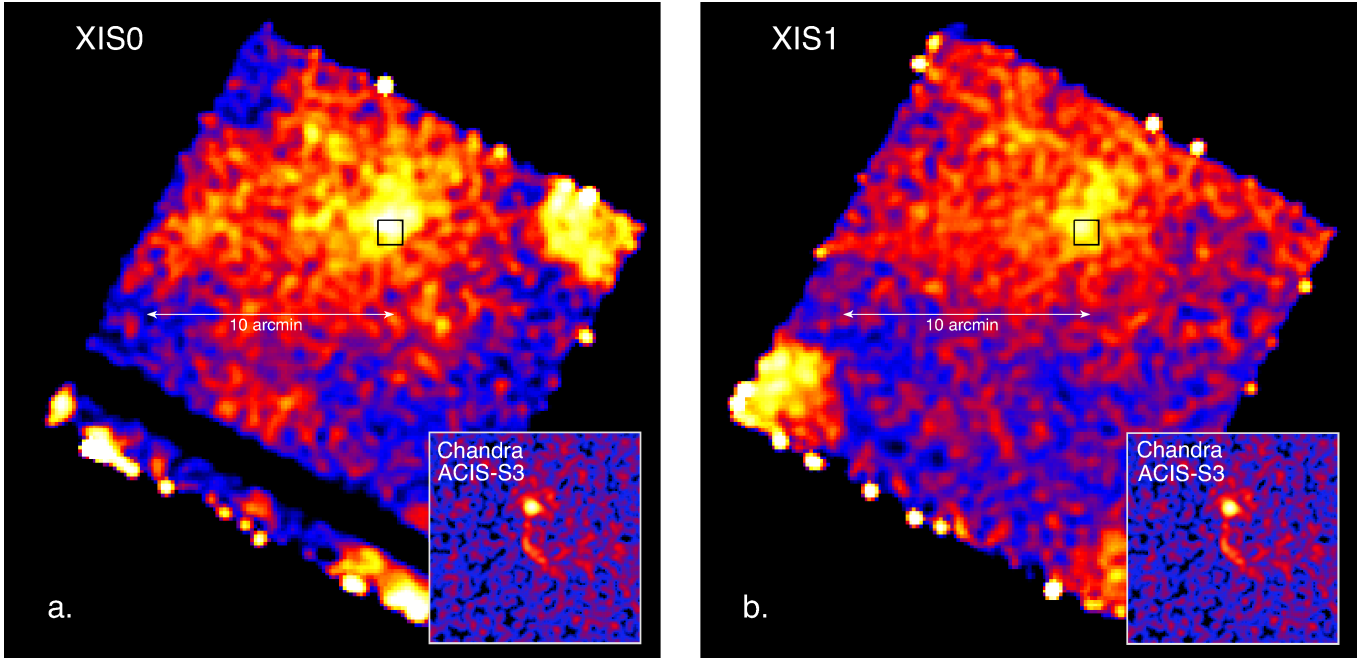}
\end{center}
\caption{\small{Vignetting-corrected images of the field around RX~J0007.0+7302 as observed by {\it Suzaku}/XIS0 ({\it{\bf panel a}}) and {\it Suzaku}/XIS1 ({\it{\bf panel b}}). A PWN component with an extent of $\sim 10'$ is revealed. 
The lost of the row image due to a putative micrometeorite hit can also clearly be seen in {\it Suzaku}/XIS0.
The inset shows the close-up view of the $1^{'}\times1^{'}$ region (illustrated by the black square) around the pulsar as seen by {\it Chandra} (adopted from \citealt{HGCH2004}). For all images, top is north and left is east.}
}
\label{pwn_img}
\end{figure*}

\subsection{XIS0 \& XIS1}
The XIS0 and XIS1 observations of 105 ks exposures were divided into two editing modes of $3\times 3$ and $5\times 5$ with $\sim 85$ and $\sim 20$~ks, respectively.
Because each pixel on the CCD was read out every 8 s in the normal mode, the time resolution of the XIS0 and XIS1 data is not accurate enough to search the periodic signal in the range less than one second.  
However, the XIS0 and XIS1 data are still adequate for the spectral analyses of the X-ray point source and the PWN to compare with the results yielded by other X-ray missions.
The adoption for the center of the point source is referred to the aim point of the {\it Suzaku} observation.
Because RX J0007.0+7302 is seriously contaminated by the surrounding PWN, we extracted the spectra of RX~J0007.0+7302 only within 1$'$.5 central circular region instead of the regular choice of 3$'$ circular region for a point source.
This adoption still contains $\sim 75\%$ of the encircled energy fraction (EEF).
We extracted the background of the pulsar from a concentric annulus of radii 100$''< r \leq 180''$ centered at the aim point of this {\it Suzaku} observation.
The spectra extracted from the point source comprise 695 counts for XIS0 and 401 counts for XIS1 after subtracting the background.
We also considered the point source selection in two other different sizes with 2$'$ and 3$'$ in radii using the same center as shown in Table~\ref{XISsp}.
Such selections of the point source contain $\sim 85\%$ and $\sim 95\%$ of the EEF, respectively.
The background of these selected regions were determined within a concentric annulus of radii 2$'< r \leq 4'$, and 3$'< r \leq 4'.5$. 
Different choices for a point source region are used to verify that the significant excess contributed by the surrounding PWN can be detected with the increasing size.

\begin{figure}
\begin{center}
\includegraphics[width=9.0cm]{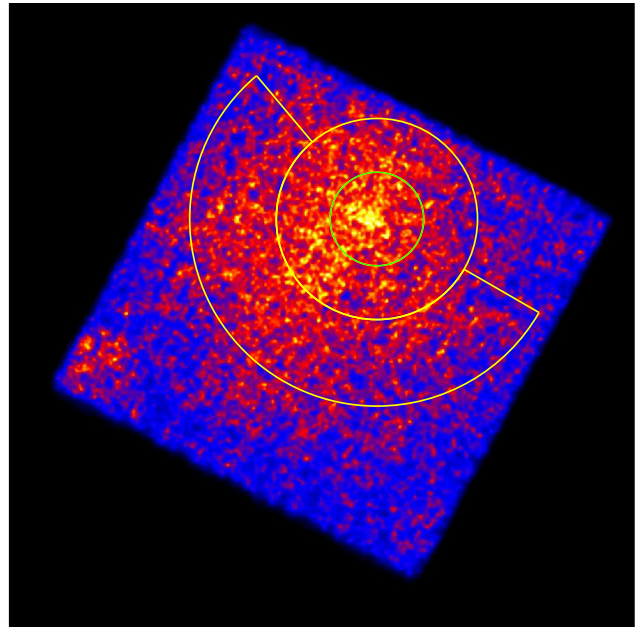}
\end{center}
\caption{\small{Illustration of the regions used to extract the spectra from different parts of the PWN associated with RX~J0007.0+7302.}}
\label{pwn_reg}
\end{figure}

For the PWN, we firstly examined its source extent by applying adaptive smoothing to both XIS0 and XIS1 images.
The images shown in Fig.~\ref{pwn_img} were generated with taking into account the vignetting of the XRT.
We used the task ``xissim''  to simulate a flat field image for our purpose, and we displayed the obtained image after dividing by the flat field. 
A large structure with an extent of $\sim 10'$ appears to extend eastward from the pulsar.
In view of the possible spectral steepening due to synchrotron cooling, we performed a spatially-resolved spectral analysis of the PWN. 
The extraction regions are illustrated in Fig.~\ref{pwn_reg}. 
We split the PWN into three regions, namely the inner part (a circular region within a radius of $2'$), the middle part (an annulus with radii 2$'< r < 4'$) and the outer part (a panda region with radii 4$'< r < 8'$).
The background spectra are sampled from the source free low-count regions from each detector respectively.
We only consider the spectra obtained with the $3\times 3$ mode to perform the spectral analysis of the PWN because of its better photon statistics.
To yield better photon statistics to describe the spectral behaviour for the PWN, we have rebinned all the spectra before the analysis. 
We generated the spectrum with $> 200$ cts/bin for the inner part, and with $> 500$ cts/bin for the middle and outer parts.

We generated the associated response matrix (rmf) and auxiliary response (arf) files by the HEASOFT command of `XISRMFGEN' and `XISSIMARFGEN'. 
The spectrum obtained from each detector was rebinned with a minimum of 30 counts per channel to ensure $\chi^2$ statistics.
We then introduced a constant in a simultaneous fit to account for the cross-calibration mismatch between different detectors.  
We also considered different circular sizes to describe the contribution of the point source.
Details about the spectral results are presented in $\S$ 3.    

\subsection{XIS3}
The XIS3 data was investigated by the timing/P-SUM mode with a long duration of $\sim 212$~ks and its image was compressed into one dimension without the information along Y-axis to obtain higher time resolution.
Our procedures to resolve the point source in a one-dimensional image follow the standard recipe for reducing XIS data with the timing mode \citep{MHT2010}.
We also redid the event reprocessing to consider the grade filtering and generated the new cleaned event file using the script `XISREPRO.XCO' with the proper selection criteria.
For the P-SUM mode, we need to remove the hot pixels on the image manually.
To identify those hot pixels, we accumulated all the photon events along the coordinate of ACTX.
We first removed those columns that have photons significantly more than others, and then we counted the distribution of events in the remaining columns and removed those pixels with the photons exceeding 3$\sigma$ than the median of the distribution.
     
\begin{figure}
\begin{center}
\includegraphics[width=8.0cm]{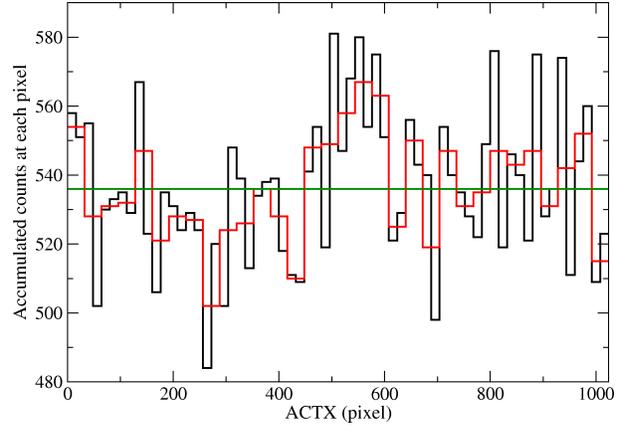}
\end{center}
\caption{\small{Histogram of ACTX versus photons distribution of the XIS3 data. The black solid line marks the average photons of each pixel after rebinning the ACTX with 16 pixels, and the red one shows the distribution with the bin of 32 pixels. The green solid line is set as 536 to demonstrate the median of the photons distribution. A point-like source can marginally be resolved within the range of 496th-608th pixel in the red line.}}
\label{XIS3}
\end{figure}

After excluding all the hot pixels, the source is still too marginal to be resolved in the histogram of ACTX versus photon events of the XIS3 data.
We then tried to rebin the distribution of counts along ACTX with several pixels as shown in Fig.~\ref{XIS3}.
A point-like source with 112 pixels ($\sim 118''$) can be roughly detected close to the center when we rebinned with 32 pixels.
There are 58255 photons for this source within 0.2-12 keV and the significance of this detection is $\sim 3.3 \sigma$.
We also checked the absolute arrival time of the photon events and determined the time required for readout by the ACTX position of the source in the XIS1 image.
In order to perform a precise temporal analysis, we corrected the arrival time with the decrease of 31.2 ms for all the source photons. 
The Solar system barycentric time correction was then implemented with the task `AEBARYCEN' at (J2000) R.A.=$00^h07^m01^s.56$, decl.=$+73^{\circ}03'08''.1$ \citep{HGCH2004} to yield a time list, which can be used to examine the pulsation of RX~J0007.0+7302/PSR~J0007+7303 in the soft X-ray band.

\subsection{HXD}
The HXD observation was conducted in nominal pointing. 
Since the HXD is a non-imaging instrument, we can not separate the contribution of the pulsar from the surrounding nebula.
However, a spectrum can be acquired to describe the behaviour of the total hard X-rays emitted from the entire CTA 1. 

We only focus on the HXD-PIN detections of the hard X-rays emitted from/around CTA~1.
In order to examine the periodic signal of the pulsar, we also applied the Solar system barycentric time correction with the task `AEBARYCEN' at (J2000) R.A.=$00^h07^m01^s.56$, decl.=$+73^{\circ}03'08''.1$ \citep{HGCH2004} on the HXD-PIN observation.
The events in the observation were restricted according to the effective energy range (10 - 60 keV) of the PIN detector, and 26661 photons were obtained for temporal analysis after the screening of the energy. 

We also subtracted the non-X-ray background (NXB) and cosmic X-ray background (CXB) to obtain the HXD-PIN spectrum. 
The simulated NXB was directly obtained from the {\it Suzaku} Data Center, and the CXB flux that is $\sim 5\%$ of the NXB for PIN could be derived from \citet{Boldt87}.
\begin{equation}\label{eqno1}
\begin{aligned}
CXB(E)=9.412\times10^{-3}\times(E/1keV)^{-1.29}\times \rm{exp}(-E/40~\rm{keV}) \\
\rm{photons~cm^{-2}~s^{-1}~FOV^{-1}~keV^{-1}}
\end{aligned}
\end{equation}
These background spectra correspond to epoch ``6'' for the flat PIN response matrix.
We also evaluated the contribution of the Galactic Ridge X-ray Emission (GRXE) using the {\it Suzaku} observation of a nearby blank sky 
(OBSID=504039010) at (l, b)=(123$^{\circ}$.9, 10$^{\circ}$.0), which is 4.3 degree apart from the CTA~1.
Following the same method as stated in \citet{Enoto2010} we assumed a power-law spectrum with a fixed photon index at 2.1 \citep{VM98}. 
The contribution of the GRXE within $14-35$ keV was evaluated to be $<4.4\times 10^{-4}$ photons sec$^{-1}$, which only corresponds
to $<$0.2\% of the NXB. 
Thus, the contribution of GRXE is negligible when we analyse the HXD-PIN spectrum of CTA~1.
After the background subtraction, we rebinned the spectrum to ensure that the photon numbers in each channel are larger than 30.

\section{Results}

\subsection{Timing Analysis}
\begin{figure}
\begin{center}
\includegraphics[width=8.0cm]{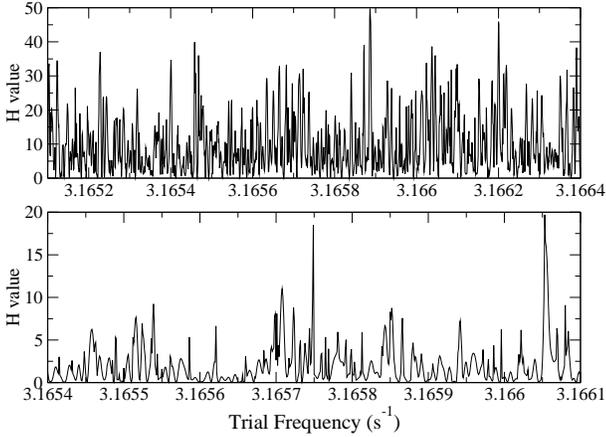}
\end{center}
\caption{\small{Periodicity search of the {\it Suzaku} data by the $H$ test. Each panel presents the search with 200 independent trials close to the expected spin frequency of PSR~J0007+7303 at the observational epoch. The upper panel shows the searching results of the {\it Suzaku}/XIS3 data, and the width of each independent trial is $\sim 4.7\times 10^{-6}$~s$^{-1}$. The lower one shows the searching results of the {\it Suzaku}/HXD-PIN data, and the width of each independent trial is $\sim 6.0\times 10^{-6}$~s$^{-1}$.}}
\label{PS}
\end{figure}

The expected spin frequency of the pulsar in our {\it Suzaku} data can be inferred from the contemporaneous {\it Fermi} ephemeris \citep{Ray2011}.
The start of GTI for our {\it Suzaku}/XIS3 and HXD-PIN data is at epoch MJD 55204.6143077.
The spin frequency of PSR~J0007+7303 corresponding to this epoch can be inferred as 3.165750271(6)~s$^{-1}$ with the first frequency derivative of $-3.6136(2)\times 10^{-12}$~s$^{-2}$ (the effect of the second frequency derivative is negligible).
We search for periodic signals with the known first frequency derivative around the predicted frequency in {\it Suzaku}/XIS3 and HXD-PIN data using $H$-test \citep{DSR89}.
The $H$-value is only 11.7 corresponding to the expected spin frequency of 3.165750271(6)~s$^{-1}$ for the {\it Suzaku}/XIS3 observation and only 6.7 for the {\it Suzaku}/HXD-PIN observation.
Fig.~\ref{PS} demonstrates the periodicity search of 200 independent trials around the expected spin frequency of PSR~J0007+7303.
We conclude no significant pulsed detection can be yielded from our {\it Suzaku} data both in the soft and hard X-ray bands. 

Comparing with the soft X-ray pulsation detected by the {\it XMM} observation \citep{Lin2010,Car2010}, our {\it Suzaku} data lose 1-dimensional imaging ability in the XIS3 timing mode and have no image ability in the {\it Suzaku}/HXD investigation.
In the {\it Suzaku}/XIS3, we can only have a marginal detection to resolve the point source with $\sim$ 58000 photons as shown in section $\S$2.2.
However, the counts gathered in the source region are much more than we expected based on the spectral behaviour provided by \citet{Sla97}. 
We noted that most of the additional counts obtained from the P-SUM mode data are caused by the instrumental background .
For the data detected in the hard X-ray band, the main contribution of the hard X-ray photons is ascribed to the X-ray background.
If we assume the pulsed spectrum of \citet{Car2010}, the expected S/N ratio of the {\it Suzaku} investigation is too low ($< 2 \sigma$) to yield the periodic signal from the pulsar.

We also examined this spin frequency using $\sim 3$-month {\it Fermi} archive (2009 November 22 - 2010 February 21) on the epochs close to our {\it Suzaku} observation.
The effective $\gamma$-ray photons of PSR~J0007+7303 were restricted in a 1-degree circle centered at  (J2000) R.A.=$1^{\circ}.7565$, decl.=$+73^{\circ}.05225$ \citep{HGCH2004} within the energy range from 100 MeV to 300 GeV.
We determined the spin frequency of the pulsar with the $H$-test using the task of `GTPSEARCH' from the {\it Fermi} Science Tools ver.~v9r15p2. 
At epoch MJD 55204.6143077, the maximum statistic was obtained at 3.16575026(1)~s$^{-1}$ with the first frequency derivative of $-3.6136(2)\times 10^{-12}$~s$^{-2}$, and this result is consistent with the aforementioned prediction as well.          

\subsection{Spectral Analysis}
\begin{figure*}
\centering
\hspace{\fill}{\includegraphics[width=6.0cm,angle=-90]{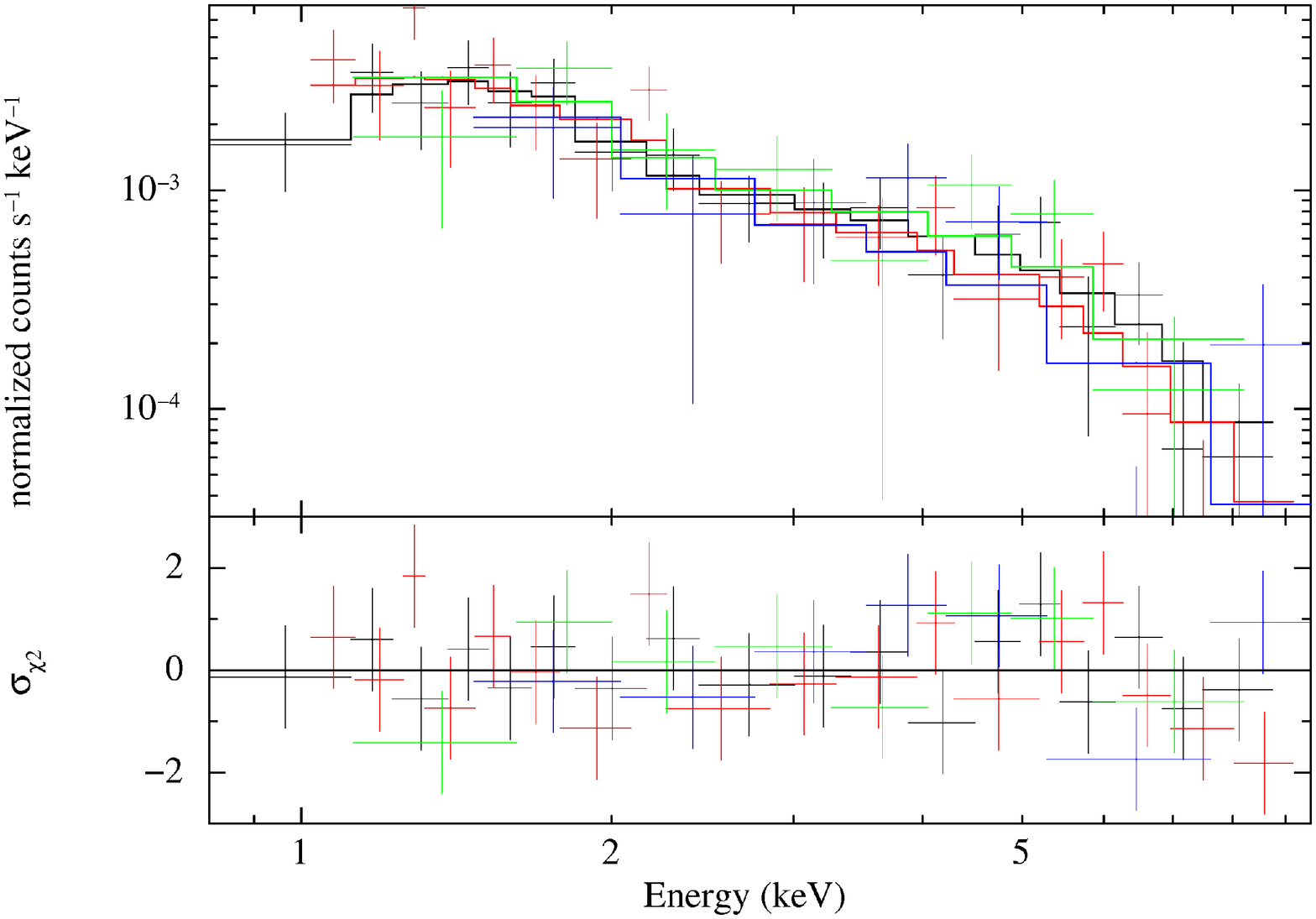}}
\hspace{\fill}{\includegraphics[width=6.0cm,angle=-90]{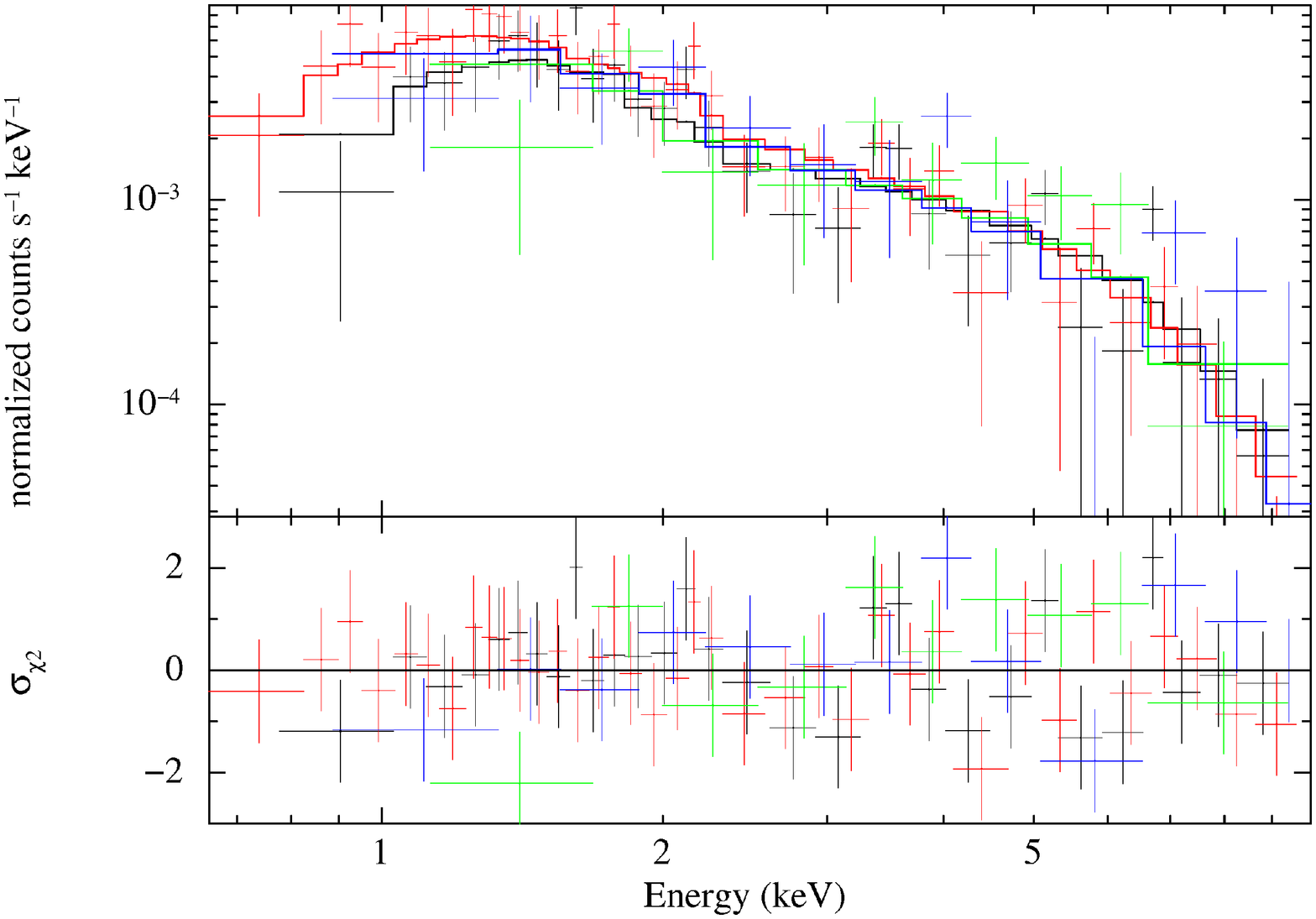}}
\hspace{\fill}{\includegraphics[width=6.0cm,angle=-90]{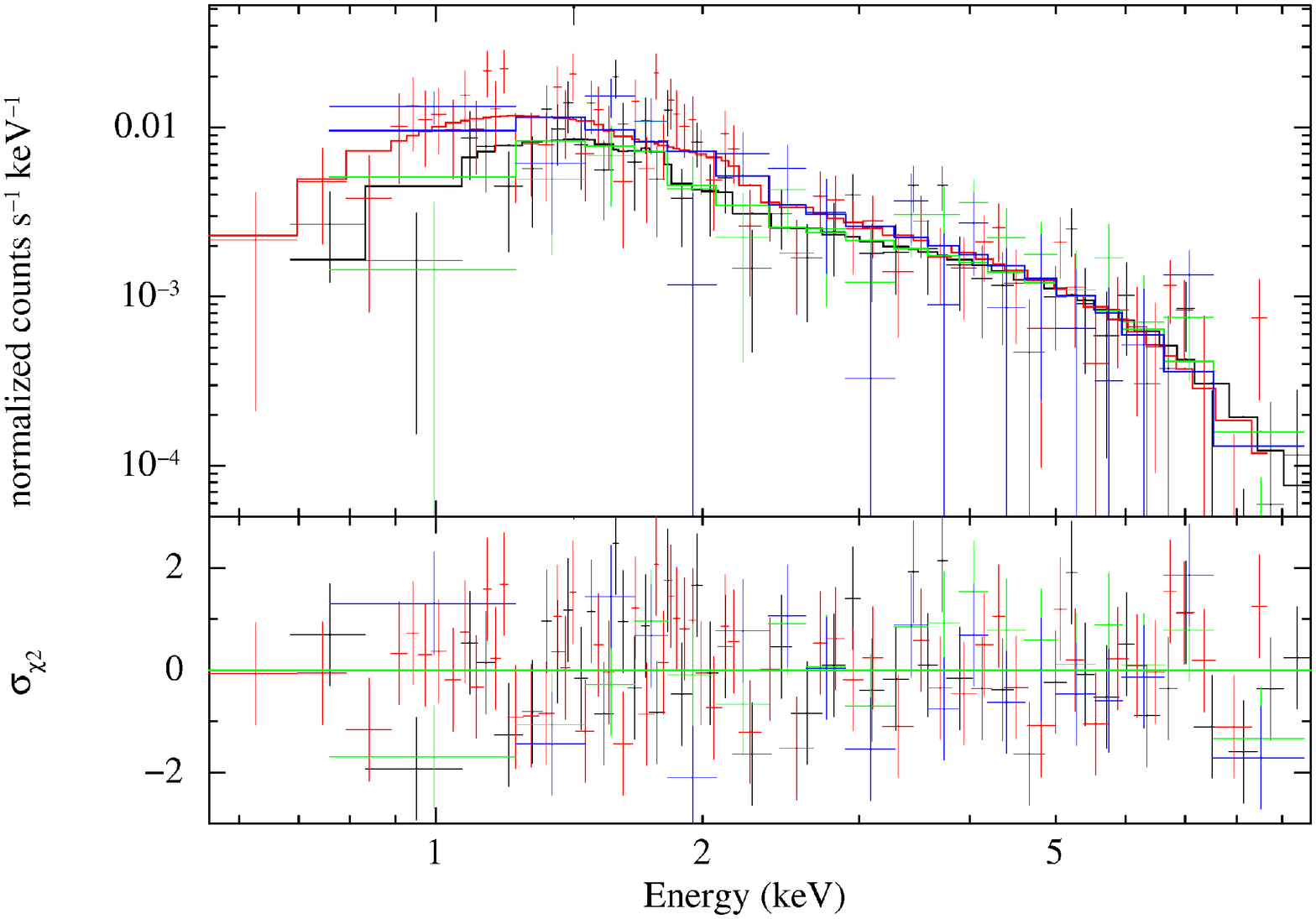}}
\hspace{\fill}
\caption{\small{Spectral fits to the power-law model. The first row presents the spectral fits to the point source centered in CTA~1 with a selected circle of 1$'$.5 and 2$'$ in radii, respectively. The second row presents the spectral fit to the point source centered in CTA~1 with a selected circle of 3$'$. The black and green data sets refer to the XIS0 observations of $3\times 3$ and $5\times 5$ modes respectively, while the red and blue data sets refer to the XIS1 observations of $3\times 3$ and $5\times 5$ modes. The detailed parameters of these fits are shown in Table~\ref{XISsp}.}}
\label{XIS}
\end{figure*}

We derived the soft X-ray spectrum from XIS0 and XIS1 onboard {\it Suzaku}.
Based on our selected region for the X-ray point source, the ratio of net source counts for the pulsar to the total photons in the source region considered in the spectrum of XIS0 is more than 25\% but in that of XIS1 is only $\sim 10$\%.
This is caused by the fact that the BI chip (XIS1) has a higher effective area at low energies.
We only considered the spectral fits in the range 0.5-10 keV and ignored those photons outside this energy range to avoid large uncertainties.
A cross-calibration term was included to correct the difference among the XIS spectra, and the result is 
consistent with the fit to the spectrum of any individual detector.
The absorption was fixed at $2.8 \times 10^{21}$~cm$^{-2}$ according to previous measurements \citep{Sla97,Lin2010}, which is consistent with optical extinction \citep{HGCH2004}.
A single power-law provides an acceptable fit as shown in Table~\ref{XISsp} and Fig.~\ref{XIS}, and our data are not good enough to give a well constraint for an additional model component, such as a thermal blackbody model or a magnetized neutron star atmosphere (NSA; \citealt{ZPS96}) model.
However, we note that the source fluxes derived from different sizes seem larger than those of previous studies and increase with source sizes; these excesses might be contributed from the PWN. 

\begin{table}
\caption{\small{Best parameters for the spectral fits to the pulsar in CTA 1 using {\it Suzaku}/XIS0 and XIS1 observations.}}
\label{XISsp}
\begin{center}
\begin{tabular}{l c c c}
\hline\hline
 & \multicolumn{3}{c}{RX~J0007.0+7302/PL} 
\\
Parameter & Small$^{b}$ & Medium$^{b}$ & Large$^{b}$ 
\\
\hline
$N_H$ (cm$^{-2}$; fixed)  & $2.8 \times 10^{21}$ & $2.8 \times 10^{21}$ & $2.8 \times 10^{21}$ 
\\
$\Gamma$ & 1.76$^{+0.27}_{-0.25}$  & 1.80$\pm$0.15 & 1.85$^{+0.14}_{-0.12}$
\\
$F_{X,\rm{PL}} \times 10^{-13}$& $3.7\pm 1.1$ & $4.6\pm 1.0$ & $7.1\pm 1.3$
\\
(ergs cm$^{-2}$ s$^{-1}$)$^{a}$ & & &
\\
$\chi^2$/dof & 39.0/46 & 78.2/85 & 157.4/143
\\
\hline
\end{tabular}
\end{center}
\footnotesize{
${}^{a}$ The unabsorbed flux is measured in the energy range of 0.3--10 keV. \\
${}^{b}$ `Small, medium and large' indicates the source selection with different size. The small, medium and large point source regions were adopted with the same center and circles with 1$'$.5, 2$'$ and 3$'$ in radii, which correspond to the EEF of $\sim$~75\%, 85\% and 95\%. The background contributions were determined within a concentric annulus of radii 100$''< r \leq 180''$, 120$''< r \leq 240''$, and 180$''< r \leq 270''$, respectively.}
\end{table}

On the other hand, our deep {\it Suzaku} observation provides a desirable opportunity to investigate the PWN, particularly for the faint emission with the extent up to $\sim 10'$ (cf. Fig.~\ref{pwn_img}). 
Previous observations with {\it Chandra} \citep{HGCH2004} and {\it XMM} \citep{Car2010} failed in detecting this large but faint diffuse structure; this is likely caused by the relatively small collecting area of {\it Chandra} and the high instrumental background of {\it XMM} respectively. 

\begin{table}
\caption{Best-fit parameters of the power-law spectral components inferred from different regions of the PWN around PSR~J0007+7303.$^{a}$}
\label{pwn_spec}
\begin{center}
\begin{tabular}{l c c c}
\hline\hline\\[-2ex]
Region & $\Gamma$ & $f_{\rm pwn}$ (0.3-10~keV) & $\chi^{2}$/d.o.f. \\
       &          & erg~cm$^{-2}$~s$^{-1}$ & \\
\hline
Inner part & $1.89\pm0.08$ & $1.2^{+0.2}_{-0.1}\times10^{-12}$ & 30.0/28 \\
Middle part & $1.79\pm0.09$ & $\left(1.4\pm0.2\right)\times10^{-11}$ & 38.1/34 \\
Outer part & $1.92\pm0.11$ & $1.5^{+0.3}_{-0.2}\times10^{-11}$ & 38.4/41 \\
\hline
\end{tabular}
\end{center}
$^{a}$ {\footnotesize{Column density is fixed at $N_{H}=2.8\times10^{21}$~cm$^{-2}$ for all these 
spectral fit. An additional power-law component to account for the pulsar emission with the parameters fixed at those inferred by \citet{Car2010} (i.e. $\Gamma=1.36$ with $f_{\rm PSR}=6.5\times10^{-14}$~erg~cm$^{-2}$~s$^{-1}$)} is included with a multiplicative factor for scaling the pulsar contributions in each region accordingly (i.e. $85\%$, $10\%$ and $5\%$ for inner, middle and outer parts respectively.}
\end{table}

In order to properly constrain the non-thermal X-ray contributions from the PWN, we also fixed the column density at $N_{H}=2.8\times10^{21}$~cm$^{-2}$. 
Besides a power-law model for PWN emission, we also included an additional power-law component to account for the X-rays from the pulsar with the parameters fixed at those inferred by \citet{Car2010}. 
For each of the three regions in our consideration, we scaled the spectral component of the pulsar with the EEF or the contribution from the wing of the PSF centered at the pulsar accordingly by multiplying a constant. 
The best-fit parameters of the power-law component of the PWN inferred from the spatially-resolved spectral analysis are summarized in Table~\ref{pwn_spec}. 
Taking the statistical uncertainties into consideration, no evidence for spectral steepening can be found from this observation. 
The whole diffuse X-ray feature can be modelled with a power-law with a photon index of $\Gamma \sim 1.8$. 

\begin{figure}
\begin{center}
\includegraphics[width=6.2cm,angle=-90]{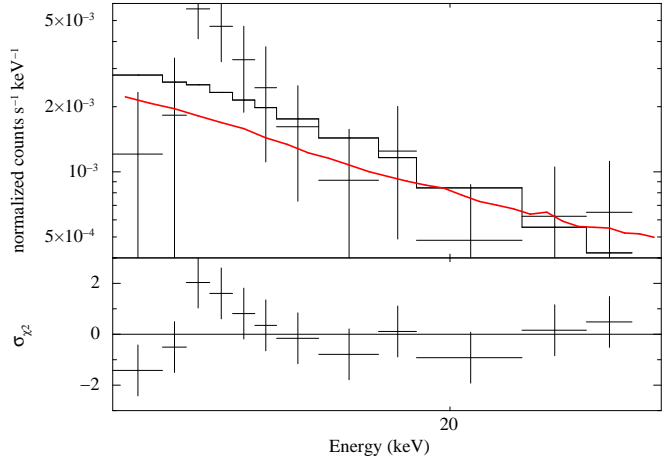}
\end{center}
\caption{\small{Spectral fit to the {\it Suzaku}/HXD-PIN observation in $14-30$ keV. The solid line represents the best-fit to a single power-law with the photon index of 3.0, and corresponds to the unabsorbed flux of  $\sim 8.0\times 10^{-12}$~ergs~cm$^{-2}$~s$^{-1}$ in 10-50 keV. The red line demonstrates the systematic errors which correspond to $\sim 5\%$ of the NXB distribution. The bottom panel shows the residuals in terms of $\sigma$.}}
\label{PIN}
\end{figure}

We also derived the hard X-ray spectrum from the HXD-PIN.
After subtracting the background, the net counts of the hard X-rays collected for the spectrum is only 6.6\% in total.
Because there are abnormal accumulations for photons at both limits of the effective energy boundary in our HXD-PIN observation, we only consider the hard X-ray spectrum within the energy range of $14-30$ keV. 
The spectrum fitted to a power-law model with the photon index ($\Gamma$) of 3.0 is shown in Fig.~\ref{PIN}.
We also checked the systematic uncertainty of the background from other observations, and the NXB accuracy should be within 5\% level in the 15-40 keV around the CTA~1 observation.
The nominal uncertainty of the adopted background model is also indicated in Fig.~\ref{PIN}.      
Because the effective data points of our hard X-ray spectrum are few, the uncertainty is very large and the best fit can not be well determined.
We fixed the absorption as 2.8$\times$10$^{-21}$~cm$^{-2}$ according to the column density obtained in \citet{Sla97}, and a photon index of $\Gamma=3.33^{+2.18}_{-1.81}$ (90\% confidence level) with $\chi^2_{\nu}=1.15$ for 10 d.o.f. can be yielded in the investigation of our hard X-ray spectrum.
Taking into account the wide spread of the acceptable photon index, the total unabsorbed flux in 10-50 keV for the whole FOV of the HXD-PIN inferred from these acceptable fits distributes from $\sim 7.8\times 10^{-12}$~ergs~cm$^{-2}$~s$^{-1}$ to $\sim 1.1\times 10^{-11}$~ergs~cm$^{-2}$~s$^{-1}$, and this signal stays between $\sim$ 3-5\% level of the NXB in the 90\% confidence level. 
The hard X-rays detected in the region around CTA~1 is also consistent with previous {\it INTEGRAL} investigation \citep{Sturn2004}. 

\section{Discussion}
We have investigated the central region of CTA~1 with a deep {\it Suzaku} observation. 
For the X-ray flux of RX~J0007.0+7302 as estimated by the XIS data, we note that it is larger than that reported by \citet{Lin2010} and \citet{Car2010} through the {\it XMM} observation.
This discrepancy can be ascribed to the fact that the PSF of {\it Suzaku} is considerably wider than that of {\it XMM}. 
In view of this, the contribution of the point source is likely to be contaminated by the surrounding PWN.
It is the main reason that we can not obtain the real spectrum of the pulsar.
Because of the loss of the imaging ability in the {\it Suzaku} data for the periodicity examination, we can not separate the instrumental effect in the soft X-ray band and the serious contamination from the X-ray background in the hard X-ray band.
These problems prevent the pulsed detection from the current {\it Suzaku} observation.
Since the {\it Suzaku} data do not provide a constrained result for the timing properties of RX~J0007.0+7302/PSR~J0007+7303, we can not obtain the phase-resolved spectroscopy and will not discuss the nature of this point source any further in the following statements.

We have also examined the properties of the PWN with the XIS data. 
A single power-law can provide a good fit and we did not detect any additional thermal X-rays emitted in the central region of CTA~1 which is consistent with previous investigation (e.g. \citealt{Sla97}). 
Both the flux and the photon index obtained in our independent investigation are fully consistent with that inferred by the recent {\it XMM} observation \citep{Car2010}. 

We would like to discuss the possible nature of the nebula that extends eastward as discovered by our deep {\it Suzaku} exposure. 
The feature revealed by {\it Suzaku} extends as large as $\sim 10'$ while the {\it XMM} observations only detected an nebular feature with an extent of a few arc-minute, which might correspond to the relative brighter component of the nebula (cf. Fig.~1 in \citealt{Car2010}). 
The spectral steepness and the flux of the inner part in our investigation are consistent with their results (refer to ``Outer PWN" in Table~1 of \citealt{Car2010}).
The morphology of this extended feature appears to be asymmetric (see Fig.~\ref{pwn_img}) which resembles those of bow-shock nebulae, e.g. PSR~J1747--2958 \citep{Gaen2004}.
Such extended feature are usually along the direction of pulsar motion and behind the bow-shock, therefore they are interpreted 
as the synchrotron radiation from the flow of particles coming out of the PWNe. 

It is instructive to discuss the plausible relation between this large feature and the torus+jet of a smaller scale reported in Figure~1 of Halpern et al. (2004; also shown in the inset of our Fig.~\ref{pwn_img}). 
The {\it Chandra} image shows that the jet have an extent of $\sim16^{''}$ toward south and bends to southwest at the far end from the pulsar (see Fig.~\ref{pwn_img}). 
On the other hand, the compact torus-like feature with a radius of $\sim3^{''}$ around the pulsar appears to be elongated in a direction perpendicular with the jet. 
\citet{HGCH2004} interpreted this system as an equatorial torus with the jet emitted along the rotation axis, which is observed in many PWN systems (cf. \citealt{NR2004} for a review). 

If the $\sim10^{'}$ feature is indeed resulted from the bow-shock as aforementioned, this suggests the pulsar is possibly moving towards west. 
This speculated pulsar motion is almost perpendicular to the portion of the jet close to the pulsar (see Fig.~\ref{pwn_img}). 
The origin of the velocity of pulsars is still an open question. 
One possible mechanism is the asymmetric supernova explosion. 
Apart from giving rise to the kick velocity, it also contributes to the initial spin of a pulsar \citep{SP98,LCC2001}. 
If the initial spin angular momentum of RX~J0007.0+7302 is dominated by the kick process, the jet emitted along the rotational axis should be more or less perpendicular to the direction of proper motion, 
which can explain the relative orientation between jet and the $\sim 10'$ feature. 
Therefore, measuring the proper motion of RX~J0007.0+7302 with multi-epoch X-ray imaging (e.g. \citealt{HB2006}) and/or dedicated $\gamma-$ray pulsar timing solution yielded from LAT data can help to constrain the relation among various extended feature of this complex system.

Despite the exact origin of this large extended feature remains to be uncertain, the X-ray spectral analysis confirms its non-thermal nature. 
Assuming it is a synchrotron nebula, we discuss its emission properties in further details. 
For the synchrotron emitting electrons distributed as $N\left(\gamma\right)\propto\gamma^{-p}$ where $\gamma$ is the Lorentz factor of the wind particles, the resultant X-ray photon index $\Gamma$ depends on whether the emission is in a fast or slow cooling regime \citep{CTW2004}. 
This is determined by cooling frequency of the emitting region $\nu_{c}=\left(18\pi e m_{e}c\right)/\left(\sigma_{\rm T}^{2}\tau_{\rm syn}^{2}B^{3}\right)$, where $m_{e}$, $\sigma_{\rm T}$, $\tau_{\rm syn}$ and $B$ are the electron mass, Thomson cross section, synchrotron lifetime and the magnetic field strength respectively. 
We have $\Gamma=\left(p+2\right)/2$ for a fast cooling scenario (i.e. $\nu_{x}>\nu_{c}$) and $\Gamma=\left(p+1\right)/2$ for slow cooling (i.e. $\nu_{x}<\nu_{c}$). 
For a standard shock model, $p$ spans a range of $\sim2-3$ (cf. \citealt{CTW2004} and references therein). 
This implies that $\Gamma$ should span the ranges of $\sim2-2.5$ and $\sim1.5-2$ for the scenario of fast cooling and slow cooling, respectively. 
Comparing these ranges with the values inferred in the XIS spectrum (cf Tab.~\ref{pwn_spec}),  we suggest that the inner and the middle parts of this PWN are probably in a slow cooling regime. 
For the outskirt of the nebula, the best-fit photon index suggests the synchrotron cooling in this region is also slow. 
However, within the $90\%$ confidence interval of the inferred photon index, it is on the margin of both regimes. 
\begin{figure}
\begin{center}
\includegraphics[width=8.0cm]{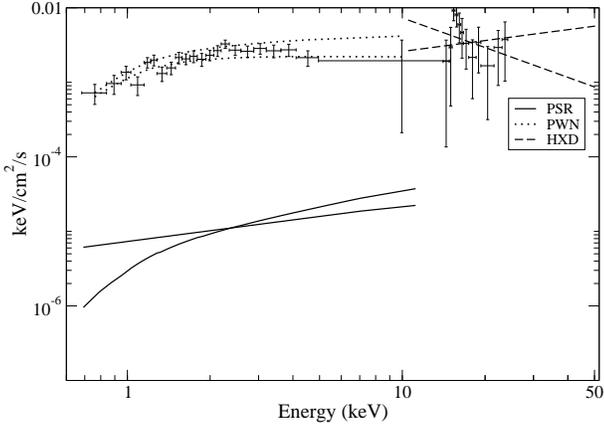}
\end{center}
\caption{Broadband absorbed spectrum of PSR~J0007+7303, PWN in the soft X-ray band and hard X-ray detection of CTA~1. The absorption in the low-energy band is fixed as $2.8\times 10^{-21}$ cm$^{-2}$ following Table~\ref{pwn_spec}. The spectral behaviour of the pulsar is determined by the power-law fit of table 1 from \citet{Car2010}. The spectrum and the data points for the PWN are derived from the outer region (the data include $\sim$ 5\% contribution of the pulsar; see Table~\ref{pwn_spec}). The solid, dotted, and dashed lines represent the uncertainty ranges of the spectral behaviour of the PSR, PWN, and HXD.}
\label{spec}
\end{figure}
For the inner and middle regions of the PWN, we speculate that $\nu_{c}$ should lie beyond the energy range covered by XIS. 
Assuming $h\nu_{c}\sim10$~keV and the synchrotron lifetime of the emitting electrons is comparable with the characteristic age of the pulsar (i.e. $\tau_{\rm syn}\sim1.4\times10^{4}$~yrs), the magnetic field strength in the emitting region can be estimated at the order of $B\sim2$~$\mu G$, which is comparable with the typical field strength in the ISM (cf. \citealt{BSSW2003} and references therein). 

Apart from the observation in the soft X-ray band, we have also investigated the field of CTA~1 with the hard X-ray data collected by HXD-PIN. 
The spectrum obtained in this hard band (i.e. $14-30$~keV) can be described by a power-law with $\Gamma=3.33^{+2.18}_{-1.81}$. 
The unabsorbed flux in $10-50$~keV is found to be $f_{x}\sim(0.8-1.1)\times10^{-11}$~erg~cm$^{-2}$~s$^{-1}$. 
We note that the photon index inferred in this band cannot be tightly constrained; however, within the $90\%$ confidence interval, the HXD spectrum can be smoothly connected with the spectrum inferred from the soft band although might have slightly spectral steepening as shown in Fig.~\ref{spec}.

We should point out that the true nature of the hard X-ray emission is not yet conclusive. 
The FOV of the HXD  is  $\sim 34'\times 34'$, while the size of PWN in the soft X-ray is only $\lesssim 10'\times 10'$.
In view of the lack of imaging capability of HXD-PIN, the observed hard X-rays can possibly be contributed by other sources in the FOV of {\it Suzaku}. This might indicate an steeper spectral break than observed.
On the other hand, the effects caused by the systematic uncertainty of the NXB background could be large. 
For example, if the NXB level is lower by a few percent (e.g. 2\%, which is the nominal value of the NXB uncertainty; \citealt{Fuk2009}) than the assumed value, the true hard X-ray would be higher than the current detection and the spectral break would not be as significant as observed. 
Since there are several upcoming missions, including {\it NuSTAR} \citep{Hailey2010} and {\it Astro-H} \citep{Taka2010}, which will be capable to image the sky up to $\sim80$~keV for the first time, the exploration of PWNe, including the one in CTA~1, will leap into a new era in the near future.

\section*{Acknowledgments}
We thank Prof. Yuji Urata in National Central University and Ms. Ting-Ni Lu in National Tsing Hua University of Taiwan for discussions on the examination of the spectra and the vignetting-correction of image for our {\it Suzaku} observation.
This work was partially supported by the National Science Council (NSC) of Taiwan through grant NSC 99-2811-M-008-057 and NSC 101-2112-M-039-001-MY3. RHHH is supported through grant NSC 99-2811-M-007-062 and NSC~100-2811-M-007-040.
CY Hwang acknowledges support from the NSC through grants NSC~99-2112-M-008-014-MY3 and NSC~99-2119-M-008-017. AKHK acknowledges support from the NSC through grant NSC~100-2628-M-007-002-MY3.
CY Hui is supported by the National Research Foundation of 
Korea through grant 2011-0023383.

 \end{document}